# ESTABLISHMENT OF THE NEW ECUADORIAN SOLAR PHYSICS PHENOMENA DIVISION

E. D. López

Observatorio Astronómico de Quito, Escuela Politécnica Nacional, Ecuador
Space Telescope Science Institute, USA

E mail (ericsson.lopez@epn.edu.ec)

*Abstract*.
Crucial physical phenomena occur in the equatorial atmosphere and ionosphere, which are currently understudied and poorly understood. Then, scientific campaigns for monitoring equatorial region are required, which will provide the data for analyzing and creating adequate models. Ecuador is located in strategic geographical position where these studies can be performed, providing data for the scientific community working for understanding the nature of these physical systems. The Quito Astronomical Observatory of National Polytechnic School is working in this direction, promoting research in Space Sciences for studying the equatorial zone. With the participation and valuable collaboration of international initiatives like AWESOME, MAGDAS, SAVNET and CALLISTO, the Quito Observatory is creating a new space physics division on the basis of the International Space Weather Initiative. In this contribution, the aforementioned initiative is presented inviting leaders from others scientific projects to deploy their instruments and to join us giving the necessary support for the creation of this new strategic research center.

**Keywords:** Space Sciences, Quito Astronomical Observatory

## Introduction

There is a relatively new field of scientific research devoted to study the physical phenomena which take place in the atmosphere in close interaction with the Sun and its variable activity. This field has been denominated Space Science, inside of which there are a lot of interesting complex phenomena currently poor understood that are waiting for sensitive instruments and adequate physical models.

Fortunately, outstanding actions like the United Nations Basic Space Science Initiative (UNBSSI) through the Committee on the Peaceful Uses of Outer Space and the United Nations Office for Outer Space Affairs (UNOOSA), provides since more than two decades ago a huge support to establish regional centers for space sciences and technology in developing countries. Moreover, the United Nations initiative has been played a pivotal role to organize the scientific community around the world through the realization of Space Science Schools, Symposiums and the annual UN workshops as those on the basis of the International Space Weather Initiative (ISWI), events that facilitate communication between space science students, engineers and scientists permitting to establish agreements for educational programs, deploying instruments in new regions and enhancing the international cooperation in research projects.

Inside the Space Sciences, one of the most interesting topics, due its general concern and direct life application, it is connected with the study of the Sun influence on the Earth climate; this is the so-called Space Weather. In this context, the United Nations Space Weather Initiative (UNSWI) have congregated leading scientists from around the world to participate of the three meeting agenda (ISWI, 2010-2012) giving continuity to the tradition of the successful International Heliophysical Year 2007 (IHY, 2005-2009). The first Workshop on ISWI was held in Helwan, Egypt and hosted by the Helwan University, Egypt, in 2010, for the benefit of nations in Western Asia. In 2011 the United Nations/Nigeria Workshop on ISWI was hosted by the Centre for Basic Space Science of the University of Nigeria at Nsukka, Nigeria, for the benefit of nations in Africa. The third ISWI workshop was hosted by Ecuador in 2012 for the region of Latin America and the Caribbean.

Ecuador has hosted the realization of the final ISWI Workshop in which key decisions have taken in order to give continuity to Space Science, technology research and education. We take advantage of this scientific meeting to promote space science studies in our country, starting with the creation of the new space station supported by the Quito Astronomical Observatory of the National Polytechnic School. The new station began with the operation of the AWESOME instrument provided inside de cooperation with the Stanford University of USA. And, with the MAGDAS instrument provided by the Kyushu University of Japan and installed at the time of realization of the UN/Ecuador Workshop.

The current paper is mainly devoted to describe the establishment of this new Ecuadorian space station named Solar Physics Phenomena as a division of the Quito Astronomical Observatory.

## The Quito Astronomical Observatory

The Quito Astronomical Observatory is one of the oldest observatories in America. This was founded in 1873 by the Ecuadorian President Gabriel Garcia Moreno as part of national government plan for investing resources for promotion and development of education, sciences and technology, at that early epoch of the republican life of the Ecuador.

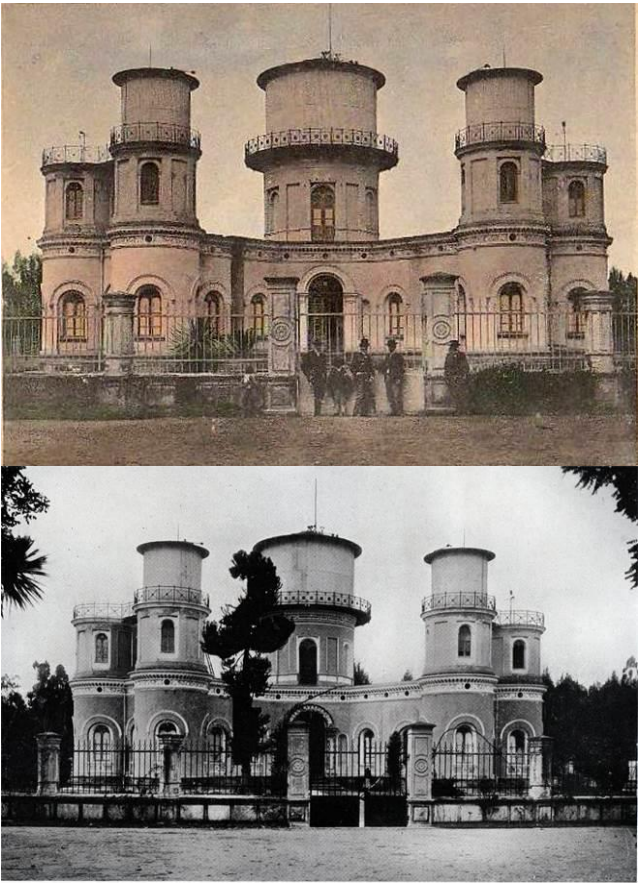

Figure 1: Photo of the Quito Astronomical Observatory (QAO), there in its early decades (upper 1899).

At the beginning the government support was exceptional providing the necessary resources that have made of this astronomical institution one of the best equipped in the world. However, after few years from its creation, the observatory suffered a long and deep lack of financial support and attention, to the point that its development and scientific production has been critical affected. This unfortunate situation, because in the country the benefits of sciences and particularly of astronomy have not been appreciated enough, has been maintained along decades being sadly reflected in the decadence and destruction of this observatory. The beautiful building was completely deteriorated, the old instruments and equipment were out of operation and abandoned, moreover all related physical and human aspects were unattended. A dark epoch of decadence, that began soon after the first decades of the observatory foundation.

In the recent time, being part of the National Polytechnic School (since 1964), almost a decade was necessary to revert this situation and recovering the Quito Observatory. The work for recovery this valuable astronomical heritage started late in 1996 and thanks to the dedicated and patient labor of several groups of students and technicians conducted by the dreams and persistence of a young Ecuadorian astronomer Director of the Quito Observatory, this noble institution currently has been completely restored in all of its aspects.

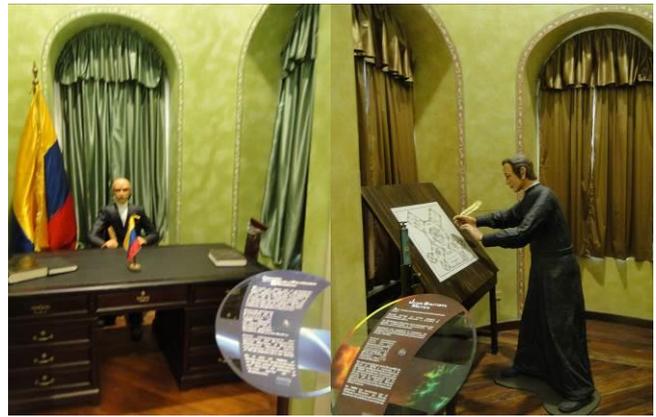

Figure 2: Garcia Moreno President of Ecuador and Founder of Quito Observatory (left). Father Juan Bautista Menten, German astronomer, designer and first Director of QAO. (right). Dummies exposed in the museum.

Today, the Quito Observatory, after a hard work, looks magnificent, in great conditions and completely operative. Actually, this is the place, where a nascent active scientific life occurs, where young students, technician and scientists all together have the serious conviction to devote their time and best effort to making of this observatory a serious and solid scientific institution contributing to understanding the physical phenomena of the Universe.

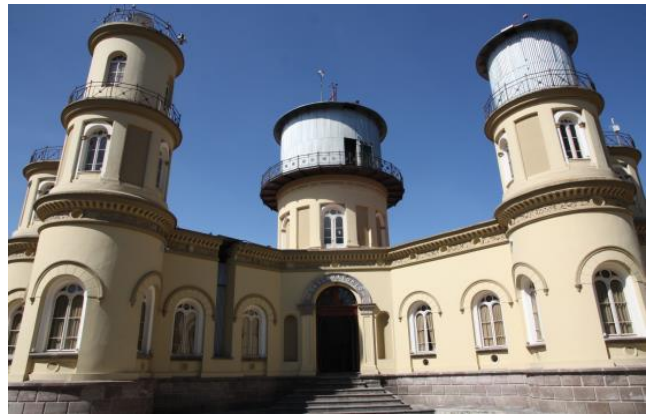

Figure 3: Building of the Quito Astronomical Observatory after the restoration in 2010.

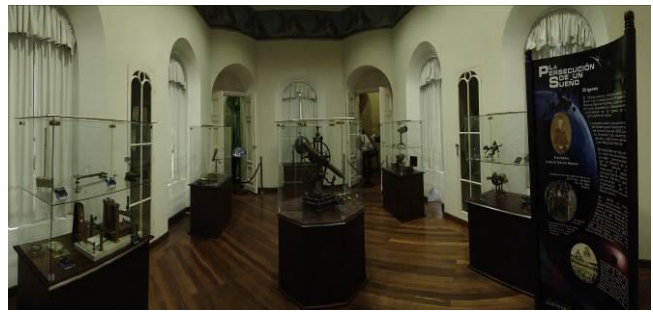

Figure 4: First QAO astronomical instruments installed in 1873.

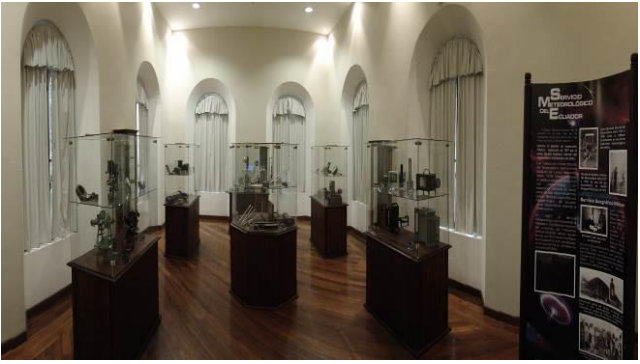

Figure 5: Meteorological instruments of the first station (1891).

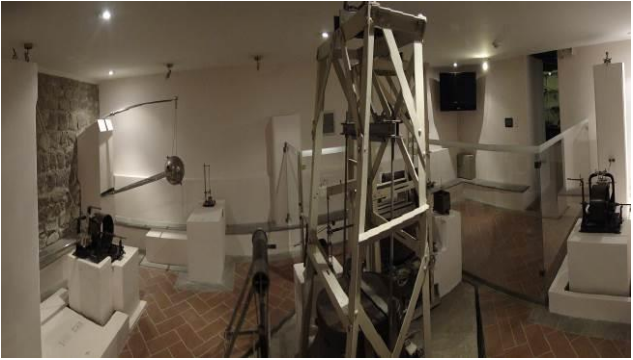

Figure 6: First seismological station in Ecuador (1904), recently restored.

Nowadays, the activities of the Quito astronomical observatory mainly are inside of three large fields: the first one is the scientific research in several areas of Astronomy, Space Sciences and Meteorology. The formal education in astronomy beside the public lectures and talks are in the second place and finally the communication with the public and outreach activities. In the following subsections we will describe in more detail these activities.

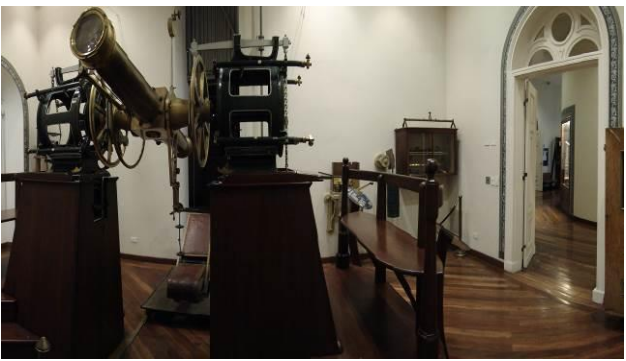

Figure 7: The Observatory Great Meridian Circle (1889)

### a) Scientific activities

The research inside the Quito Observatory has been planned to go mainly inside the frames of theory and data analysis. This aspect is in close connection with the local earth-atmospheric conditions and the economic situation of the country. In Ecuador, there is not an adequate place where an optical observatory can be established; the high frequency of cloudy skies along the year is a serious impediment and it is a natural consequence of the singular topography of the Ecuadorian lands; high mountains in the cordillera surrounded by the neighbor tropical regions of pacific coasts and the eastern amazon jungle. Beside this, there is the permanent difficulty to raise funds from the government, necessary resources to provide adequate observational facilities. Consequently, the best choice that we have is to drive the scientific institutional activities toward modeling and theoretical work. However, the observational necessities mainly linked with the special geographical position of the country on the equator, are considerate under a future project for the construction of a radio astronomical observatory, designed in such way to avoid the climate inconveniences.

The theoretical work, data processing and analysis have started with some interesting contributions inside of High Energy Astrophysics (Active Galactic Nuclei, Quasars and microquasars, Gamma Ray Burst) and cosmology (Models with non-zero cosmological constant, microlensing and standard candles). Forward, we are looking to incorporate new branches as the radio astronomy and the studies in the infrared region of the electromagnetic spectrum.

The main idea behind the institutional plan is to organize scientific groups around well-prepared leading researchers with a large expertise and with the capability to gathering students. In this way, we expect to grow quickly and strongly with an appropriated competitive scientific level.

Thanks to the United Nations contribution promoting the space sciences research, the Quito Astronomical Observatory has appreciated the importance to include in its institutional plan of scientific interests, the space science research. Ecuador, located in a particular geographical position, is potentially interesting to install instruments from the International Space Science community, in order to provide a new source of valuable scientific data for the study of space weather influences on the Earth atmosphere. This will bring a great benefit to the scientific community providing valuable information to understanding ionospheric and magnetospheric phenomena which are taking place in the equatorial zone.

### b) Education in astronomy and space science.

In Ecuador, formal education in astronomy unfortunately is not provide at schools, high schools even not at the university level. In the pass, we did several unsuccessful attempts to incorporate subjects of astronomy in the student curricula, but the administration of the initial and middle education in the country did not give the necessary importance to these topics. We hope for new opportunities to maintain conversations with local educational authorities, with the aim to include adequately in the program of studies subjects touching the education in astronomy and space sciences.

In these circumstances, currently the only place that we have to provide formal education in astronomy is at the university level and only in some institutions where there is this interest. In this context, the Quito Astronomical

Observatory plays a pivotal role for promoting the education in astronomy and space sciences in Ecuador.

The Escuela Politecnica Nacional is one of few institutions where the attention to the education in astronomy is devoted. This is thanks the existence of the Sciences Faculty and its Physics Department, where the students are formed to do research in several branches of physics and among these in astronomy and astrophysics. The professional astronomers that Ecuador has initially have been formed in this department and later we have completed our careers with postgraduate programs in astronomy in several countries abroad. The Quito Observatory is involved in these educational activities of the physics department and the education in astronomy has been provided by its staff. Currently, we are planning a more aggressive program to intensify the formal education in astronomy at the Polytechnic School and to promote the expansion to other academic institutions.

Beside the formal education in the classroom, scientific meetings, technical and academic talks, stargazing sessions, are organized to stimulate the interest and to spread ideas among the students and professional working in astronomy.

### c) Promotion of astronomy and communication with the public

This is the third natural activity that the Quito Observatory has. Locally, news, data and reports on astronomy are provided by the Quito Astronomical Observatory. For the media this Observatory is the primary source of astronomical information.

On the other hand, in order to involve the community and in particular children and students with the sciences, the Observatory offers several facilities: the first Ecuadorian astronomical museum has been created on the basis of the old instruments and equipment of the Quito Observatory heritage. This new museum located inside of Observatory installations in the Alameda Park in Quito downtown was totally open in the summer of 2012. A large collection of astronomical, meteorological and seismological instruments are in permanent exhibition for the public.

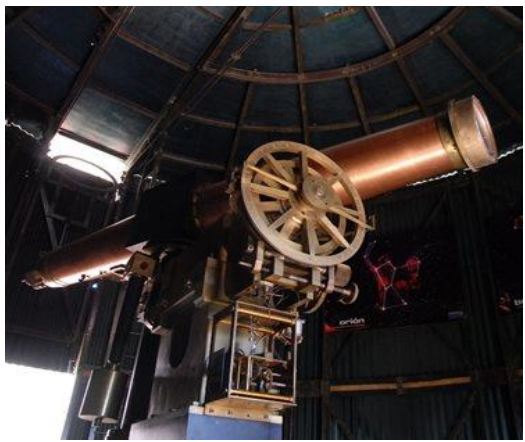

Figure 8: Ecuatorial Merz telescope made in Germany (1875)

Otherwise, during the clear skies in Quito, the Observatory opens its door to the public to spend nights in stargazing sessions. Unfortunately the quantity of clear nights in Quito and in general in Ecuador is low and often the best time to organize these observation campaigns, using the telescopes available in the Observatory, is during the summer season. These astronomical observations and specially videos of astronomical event like eclipses, comets, conjunctions, etc. are accessible to the entire population online. The telescope views are broadcasted by the internet and anyone has the possibility to participate in live of these observations no matter where are located. This is the streaming astronomical system of the Quito Observatory. Currently, this system that uses two large telescopes has been disabled due the update and improvement that the technical personnel of the Observatory do.

Finally, from time to time, the Observatory organize astronomic traveling fairs visiting several cities of Ecuador, engaging the population and local schools, in a program of astronomy with instruments exhibitions, talks, games and night observations.

Beside these outreach activities, the periodic publications of the Observatory contribute to popularization and divulgation of astronomy and space sciences.

## Solar Physics Phenomena Division

Thanks to the valuable support of the United Nations Office for Outer Space Affairs, many schools and workshops on the basis of the International Space Weather Initiative have been realized around the world, providing the opportunity to instrument deployment and the close international scientific collaboration. In particular, as result of this important cooperation, in the recent years Ecuador has been involved in space sciences research and its leading institutions are deeply commitment and decide to dedicate efforts and resources to contribute jointly to study the physical phenomena which occur in the equatorial region of the atmosphere, in close interaction with the Sun. As part of this interest with the support of the Ecuadorian government, the United Nations, NASA among other important contributors, the UN/Ecuador Workshop hosted by the Quito Astronomical Observatory of the Escuela Politecnica Nacional was held in Quito from 8 to 12 October 2012. This was the twentieth in a series of Workshops on basic space science, the International Heliophysical Year 2007 and the International Space Weather Initiative,

This workshop was mainly focused on space sciences and its main objective was to provide a forum in which participants could comprehensively review achievements of the International Space Weather Initiative and further plans for the Initiative, as well as assess recent scientific and technical results in the field of solar-terrestrial interaction.

The 2012 UN/Ecuador Workshop was a huge real contribution that has established the primary base for incorporating Space Sciences studies in Ecuador, enforcing the parallel effort that scientist and engineers of neighbor countries, like Peru and Colombia do in the same context.

As a wonderful consequence of this ISW Workshop was the agreement and support of our authorities and in particular of the National Polytechnic School, who understanding the importance to carry out space sciences studies in Ecuador have demonstrated strong interest and the compromise to give the necessary support for the development of these studies. This commitment starts with the acceptance for the creation of the Solar Physics Phenomena Division of the Quito Astronomical Observatory, which will be in charge and with the responsibility to promote and realize studies of physical phenomena that take place in the equatorial region liked to the influence Sun on the Earth atmosphere.

Currently, in the nascent space station two instruments are in operation: the Atmospheric Weather Electromagnetic System for Observation Modeling and Education (AWESOME) system, installed in 2010 under the collaboration of USA (Dr. Umran Inan, scientific leader) and the Magnetic Data Acquisition System (MAGDAS) installed at the time of the UN/Ecuador Workshop, with the cooperation of Japan (Prof. K. Yumoto, and G. Maeda scientific leaders). Both instruments are in operation and the calibration process is going on to guarantee the quality of the data.

Other important collaborations are coming: the South American VLF NETwork (SAVNET) currently in implementation, the solar spectrometer Compound Astronomical Low-cost Low-frequency Instrument (CALLISTO), the Dual frequency GPS Network, are expected to join the near future the bunch of instruments installed in Ecuador. The respective conversations are in process, discussing the feasibility of these projects and their scientific benefit and contribution to the space science studies.

In this context, the new Solar Physics Phenomena station of the Quito Astronomical Observatory initially will be structured with 5 instruments that will provide unique data for the study of the space weather influences on Earth`s atmosphere. The data sets of this station will be distributed and publicly available to guarantee the mayor utility and applicability. The main objective is to create a well-established space station, operating for the benefit of the local, regional and in general for the global scientific community around the world, who will be able to obtain quality data for space science studies of the equatorial atmosphere.

As a magnify result of the UN/Ecuador Workshop, the Rector of Escuela Politecnica Nacional of Ecuador has been committed with the project for the creation of Quito Solar Physics Phenomena Division and actually, as the maximum authority of the University, he kindly is providing the necessary support for the construction of the new building for the space station, work that is going on. On the other hand, students from the Sciences Faculty has been interested in space science topics and after the Workshop they have joined the Quito Observatory staff, and they are working in learning the physics of atmosphere in interaction with the Sun and understanding the processes related to data collection and processing technics. We hope that the first scientific papers will yield soon derived from the ionosphere and magnetic field data from the AWESOME and MAGDAS instruments, currently in operation.

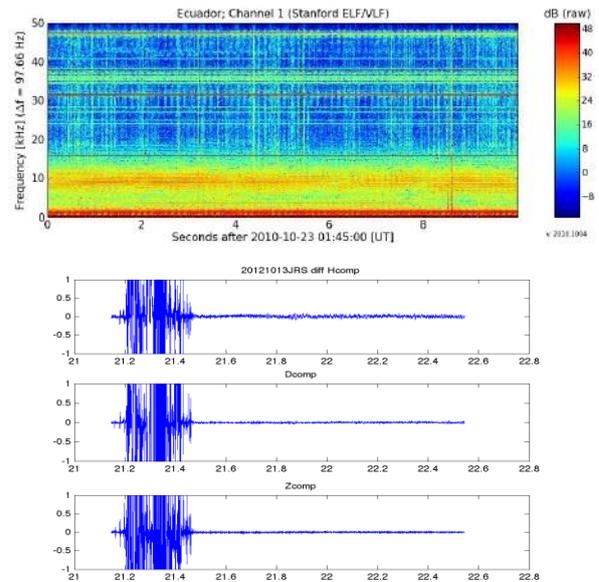

Figure 9: Examples of the first spectra from the AWESOME (up) and MAGDAS (down) instruments installed in Ecuador

## 2012 UN/Ecuador Workshop on ISWI

Without doubt a great event in the pass year was the realization of the 2012 UN/Ecuador Workshop on the International Space Weather Initiative. This was possible mainly thanks the enormous support from the United Nation Office for the Outer Space Affairs (UNOOSA), the Secretary of Higher Education, Sciences, Technology and Innovation of Ecuador (SENESCYT), the National Aeronautics and Space Administration of USA (NASA), the Quito Astronomical Observatory of National Polytechnic School of Ecuador (OAQ-EPN), among other important contributors and supporters. This scientific meeting was attended by about 100 participants from 20 countries, with oral and poster contributions inside the Workshop topics: Observations of the Sun, Ionosphere and Magnetosphere, Very-Low Frequency (VLF) studies of Sun-Earth Connection, Climate Studies, Atmospheric Physics, Space Weather Modeling and UNBSSI Follow-up Projects in Astronomy.

The UN/Ecuador Workshop has importance for Ecuador in order to promote and stimulate space science studies in this equatorial region, where no research in these topics has been done before. Now, under the umbrella of the Quito Astronomical Observatory and with the support of the scientific community we have the opportunity to contribute to understanding the physical process behind phenomena which take place in the complex Solar-Earth system. For that, scientists and students are being involved in operating and processing data obtained from instruments installed with the support of international research groups, for study the equatorial atmosphere.

We conclude that the UN/Ecuador Workshop, the third and final of the International Space Weather Initiative series, was very successful and brought a lot of new experiences and benefits to each participant. Particularly, in the case of Ecuador, we can summarize the following important outcomes of the Workshop: first, actually Ecuador is measuring the local magnetic field in the Jerusalem place, where the magnetometer of the MADAS array was installed. This was possible thanks to the valuable contribution of the Kyushu University of Japan. The Workshop was essential to consolidate the local support of our authorities for the creation of the new space sciences station in Ecuador, to which future collaborations from the nets: Savnet, Gps network, Callisto will join. On the other hand, possible new space sciences events in Ecuador are in mind for the realization in the next future, as the 2014 space science school on MAGDAS data processing. Beside that new students are interested and involved in space science studies. New contacts and future projects in joint research collaboration have been coordinated during the Workshop. However, from our point of view, a relevant result of the meeting is the fact that space science studies are better understood and appreciated by both the local government and university authorities. This point in turn could facilitate the development of space science studies providing the required financial support. Finally, the regional interest for mutual international cooperation and the trust on Ecuador capability to carry out space science studies have been enhanced.

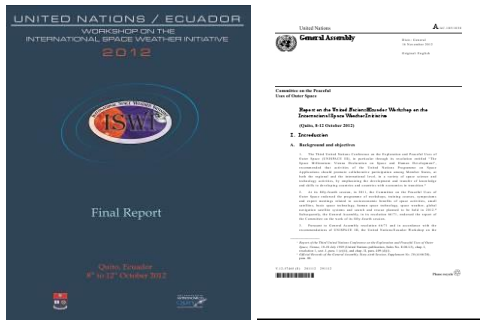

Figure 10: Resolutions of the UN-Ecuador Workshop on the International Space Weather Initiative October 12, 2012, Quito, Ecuador.

The UN/Ecuador Workshop had key importance for taking decisions to guarantee the support and promotion of Space Sciences studies around the Globe. In this context, essential recommendation has been resolved during the five-days meeting, resolutions incorporated in similar reports issued by the United Nation Office for Outer Space Affairs (UNOOSA) and the Quito Astronomical Observatory (QAO). We considered important to bring below the text of some of these recommendations with the purpose to assist in their dissemination:
- It is recommended that the ISWI continue the operation and development of existing arrays and deployment of new instrument arrays as appropriate.
- It is recommended that the ISWI undertake a process to examine data sets to determine data utility, to develop connections with virtual observatories to make data more readily available, and to facilitate collaborative modeling of regions of interest (e.g. the equatorial ionosphere) in collaboration with modeling centers of the ESA, JAXA, NASA, and others.
- It is recommended that data from ISWI instrument arrays be combined with space-based and other ground-based data to advance space weather science leading to robust research output and scientific papers in international journals. It is recommended that ISWI and GNSS communities collaborate in terms of data sharing and space weather research.
- It is recommended that the ISWI Space Science Schools and the annual UN workshops for ISWI continue indefinitely. UN/BSS workshops and Space Science Schools are an integral part of ISWI, to train early career and new researchers in instrument operation and the science of heliophysics. The partnerships already established with international scientific organizations need to be strengthened to assure that these capacity building activities are accomplished efficiently and for the benefit of all member states.
- It is recommended that new knowledge generated by ISWI activities be effectively communicated to the public and the scientific community at large via Newsletters, ISWI web site, and other media.

Given the enormous contribution that Japan has made to the astronomy and space science communities, it is not impossible for Japan to continue the operation of the Nobeyama radioheliograph on a long-term basis. The international scientific community will be grateful if the Nobeyama radioheliograph is made to survive and the effort will be recorded as another outstanding Japanese contribution to the humankind. Therefore, the participants of the UN/Ecuador workshop on International Space Weather Initiative strongly recommend the continued operation of the Nobeyama Radioheliograph either by the current institution or by a consortium of new institutions.

## Conclusive Remarks

Since the Space Sciences studies and in particular the space weather is an international matter, a mutual effort from all nations should be done in order to promote the deployment of instruments in regions unobserved before, and to guarantee the continuity of high quality data acquisition, processing and modeling.

The International Heliphysical Year 2007 and International Space Weather Initiative made significant contributions in the installation of new instrumentation, and also providing the opportunity to meet scientists, technicians working in space sciences, creating with the Space Workshops and Schools appropriate spaces where is possible to exchange ideas, review new achievements, plan future activities and discuss mutual collaboration in specific projects.

Considering that the United Nations and Space Agencies contribution have been huge effective and determinant for the development of Space Sciences studies around the World, it is fundamental that under this support will continue, as part of the Space Weather agenda item of the Scientific and Technical Subcommittee of the Committee on the Peaceful Uses of Outer Space, in 2013 and beyond.

The United Nations and Space Agencies initiative on the peaceful uses of outer space and space weather studies plays a great role as the irreplaceable component for promoting and organizing scientific studies around the world for understanding the behavior of the Sun and its influence in the Earth-atmosphere and the global climate. Without this organizing component, the efforts could be individual, isolated and in certain sense disordered provoking a slow growing and development of the scientific activity in the space sciences field with the consequent misunderstanding or poor comprehension of the physical phenomena that take place in the Sun-earth-atmosphere system. Fortunately this is not the case and the research activities in this field have been greatly promoted and boosted by the Space Sciences Initiative of the United Nations. Consequently, at the level of development of the current space science structures, the support of the UN and space agencies is fundamental and the priority should be to maintain these studies for the benefit of the all nations.

The development of sciences in great part depends on the available facilities and its adequate organization. On the other hand, each nation and its researchers at the same time should contribute with the local support for organizing space science centers, training students and using the data for the corresponding studies and production of scientific papers. In this context, Ecuador through the Quito Astronomical Observatory of the National Polytechnic School have been inserted in the space science process and with the Japan international cooperation of MAGDAS, a space weather instrument have been installed in Jerusalem site located a 20 Km northward from Quito, this is a valuable and an immediate result of the UN/Ecuador Workshop. Moreover, the solar physics phenomena station is being created and the Quito Astronomical Observatory of the Escuela Politecnica Nacional of Ecuador had offered to act as a regional center for space weather science and education.

In the meantime, we have the intention to organize a Space sciences school on MAGDAS data analysis and modeling, in joint collaboration with Japan. In the same way, we are ready to perform the necessary work to introduce the space science and technology education at the elementary, secondary and university level together with the astronomy.
We expect that this initial work inside of the Solar Physics Phenomena Division of the Quito Astronomical Observatory will be the basis to promote and strengthen space science studies in Ecuador, where a significant number of high level researchers and students could be potentially involved.

## Acknowledgments
EL gratefully acknowledges the continuous support from the United Nations, and the space agencies: NASA, JAXA, ESA, for development space science studies around the world and for making possible the realization of the 2012 UN/Ecuador Workshop for benefit of South America and the Caribbean. Similarly, he wants to thank and recognize the huge and permanent contribution from Dr. Sharafa Gadimova; her valuable work always was a real guarantee for the realization of this Workshop.
Special thanks to Prof. Yumoto and George Maeda for the installation of the MADAS instrument in Ecuador near at the time of the UN/Ecuador Workshop. My deep gratitude to Ecuadorian government Secretary Senescyt and its personnel for the valuable support and contribution to development of sciences and technology in the country. In the same spirit our recognition to the Escuela Politecnica Nacional for the continuous support offered to the Workshop and to preparing this manuscript. E.L. was supported by the National Secretary of Higher Education, Science, Technology and Innovation of Ecuador (Senescyt, fellowship 2011).

## References

- Burgos M.:2011, Observatorio Astronomico de Quito: una puerta al Universo, Arte y Cultura, http://elimperdible.ec/web/arteycultura.
- Lopez E. (Ed.): 2005, 132 años de Historia del Observatorio Astronomico de Quito, Nina Comunicaciones, Quito, Ecuador, 2005.
- UNOOSA and QAO: 2012, UN/Ecuador Workshop reports.